\documentclass[aps,preprint,amssymb,12pt,floatfix]{revtex4}
\usepackage{subfigure}
\usepackage{graphicx}
\usepackage{rotating}

\begin{document}

\title{Asymmetry in the shapes of folded and denatured states of proteins}

\author{Ruxandra I. Dima and D. Thirumalai}
\affiliation{\em \small Institute for Physical Science and Technology and\\
                Department of Chemistry and Biochemistry\\
        University of Maryland, College Park, MD 20742}

\date{\small \today}

\baselineskip = 18pt

\begin{abstract}

The asymmetry in the shapes of folded and unfolded states are probed using
two parameters, $\Delta$ (a measure of the sphericity) and $S$ that
describes the shape ($S > 0$ corresponds to prolate and $S < 0$ 
represents oblate). 
For the folded states, whose interiors are densely packed, 
the radii of gyration ($R_g$), $\Delta$, and $S$ 
are calculated using the coordinates of the
experimentally determined structures.  Although $R_g$ scales as 
$N^{1/3}$, as
expected for maximally compact structures, the distributions of $\Delta$
and $S$ show that there is considerable asymmetry in the shapes of folded
structures. The degree of asymmetry is greater for proteins that form
oligomers. Analysis of the two- and three-body contacts in the native 
structures shows that the
presence of near equal  number of contacts between backbone and 
side-chains
and between side-chains gives rise to dense packing. We suggest
that proteins with relatively large values of $\Delta$ and $S$ can
tolerate volume mutations without greatly affecting the network of
contacts or their stability.  Mutagenesis data on T4 lysozyme and
$\lambda$-repressor support this conclusion. 
 To probe shape characteristics of denatured states
we have developed a $C_{\alpha}-C_{\beta}$ model of a WW-like
domain.  The shape parameters, which are calculated using Langevin 
simulations,
change dramatically in the course of coil to globule
transition.  Comparison of the values of $\Delta$ and $S$ between the
globular state and the folded state of WW domain shows that both 
energetic (especially dispersion in the hydrophobic interactions)
and steric effects are important in determining packing in proteins.

\end{abstract}

\maketitle

\newpage

{\bf INTRODUCTION:}\\

The densities of the core of the compact folded states of 
proteins approach that observed in small-molecule
organic crystals \cite{RichardsQuartRevBioph94}. 
Based on the observation that packing densities of proteins are
high it can be argued 
that the requirement of dense packing in the core may greatly
restrict the number of sequences that can fold into a unique
structure \cite{RichardsQuartRevBioph94,RichardsJMB74,FinneyJMB75}. 
The need for folded proteins to have a high packing density is not so stringent a requirement
that volume mutations cannot be made. Indeed, 
extensive 
experiments on a number of proteins show that relatively large 
volume changes (up to about 10 methylene equivalents) can be made
without drastically affecting protein stability
\cite{RichardsQuartRevBioph94,LimBiochem92,KarpusasPNAS89,HurleyJMB92}. 
Even large
volume mutations can be accommodated by minor variations in the
secondary structure without compromising the global fold 
\cite{LeskJMB80}. However, it appears that
maintaining the hydrophobic nature of the core 
may be the key factor in establishing high packing density. 
Current experimental evidence suggests that
excluded volume effects and hydrophobic interactions are 
the major factors that give rise to high packing density 
\cite{RichardsQuartRevBioph94}. 

Recently, Liang and Dill \cite{LiangBiophysJ01} have reexamined
packing in proteins using a Delaunay triangulation method of
dividing space. This approach, which is used to compute 
the distributions of voids and occupied regions, shows that 
both high-density liquid-like behavior and solid-like
characteristics are seen in proteins. The free volume distributions
suggest that proteins have the characteristics of high-density
liquids and glasses that are often accompanied by the
presence of ``defects''. 
There may be a dual requirement on the overall structure of
proteins: (i) Marginal stability of folded state can be achieved
by having high packing density, but not necessarily close packed structures; 
(ii) Plasticity of proteins, as evidenced by their ability to tolerate
mutations, is better accommodated by non-spherical folds with voids.
Perhaps, the two requirements are
required for foldability and function of proteins.

Because proteins form dense packed structures, we expect 
that their overall shapes would be spherical.
The purpose of this paper is two fold: 
(a) We use ideas familiar in
polymer physics \cite{SolcJCP71,AronovitzJP86,HoneycuttJCP89} 
to describe shape parameters of proteins
using only the coordinates of the native structure. Examination
of the asymmetry in the overall shapes gives insights into the
extent to which a protein can tolerate volume mutations. 
(b) The folding pathways of proteins depend on the nature of the
denatured states which are difficult to characterize. To shed
light on the changes in shapes of proteins that occur in the 
course of a folding reaction we have used a simple off-lattice
C$_{\alpha}$-C$_{\beta}$ representation of a WW-domain to model
denatured states. Comparison of the shape parameters (see below)
of the folded and denatured states shows that considerable 
asymmetry is lost as the protein folds. These calculations
show that the degree of asymmetry in the folded states of
proteins arises due to dispersions in the energetic 
interactions involving side chains and the peptide backbone.

{\bf METHODS}\\

{\bf Shape parameters}

To determine the shape of a given conformation we use 
two rotationally 
invariant quantities $S$ and $\Delta$ that are determined
using the inertia tensor \cite{AronovitzJP86,HoneycuttJCP89,CannonJPhys91,JagodzinskiJPhys92},
\begin{equation} 
T_{\alpha \beta} = \frac{1}{2 N^{2}} \sum_{i,j=1}^{N} (r_{i \alpha} - r_{j \alpha}) (r_{i \beta} - r_{j \beta})
\label{eqn:T_tensor}
\end{equation}
where $N$ is the number of atoms in the protein, $r_{i \alpha}$ is the 
$\alpha$-th component of the position of atom $i$ and
$\alpha, \beta$ = x,y,z. The eigenvalues of {\bf $T$}, denoted by 
$\lambda _{i}$, 
are the squares of the three principal radii of gyration. 
Thus, $R_{g}^{2} = trT = \sum_{i=1}^{3} \lambda_{i}$.

Asphericity is measured using

\begin{equation}
\Delta = \frac{3}{2} \frac{[\sum_{i=1}^{3} (\lambda _{i} - \overline{\lambda})^{2}]}{(tr {\bf T})^{2}}
\label{eqn:Asphericity_def}
\end{equation} 

\noindent where $\overline{\lambda} = \frac{tr {\bf T}}{3}$ is the average eigenvalue
of the inertia tensor. Deviation of $\Delta$ from 0 (the value 
corresponding to a sphere) gives an indication of the extent of
anisotropy. The overall shape of a protein is characterized using

\begin{equation}
S = 27 \frac{[\prod_{i=1}^{3} (\lambda _{i} - \overline{\lambda})]}{(tr {\bf T})^{3}}
\label{eqn:Shape_def}
\end{equation} 

\noindent
Negative
values of $S$ correspond to oblate shapes and positive values to prolate 
shapes. For perfect spheres $S$ and $\Delta$ are: $S$ = $\Delta$ = 0. 
The parameters $\Delta$ and $S$ obey the inequalities
$0 \leq \Delta \leq 1 $ and
$-\frac{1}{4} \leq S \leq 2$.

The advantage of quantifying shapes in 
terms of $\Delta$ and $S$ is that they can be directly computed using
{\em only the coordinates of} the experimentally determined structures without
assumptions about overlaps between van der Waals volumes assigned to
atoms \cite{LiangBiophysJ01} that are used in the Voronoi analysis 
\cite{RichardsJMB74,FinneyJMB75}. 
The shape parameters are directly calculated from the
inertia tensor which uses only the structures of proteins.
Because the most dense packing in the 
native state occurs for $\Delta$ = $S$ = 0 it follows that deviations of
$\Delta$ and $S$ from zero would suggest that packing is 
compromised.

To obtain the distribution of $\Delta$ and $S$ for folded proteins,
we chose a representative set of protein structures from the Protein Databank 
(PDB) \cite{BermanNAR00},

\noindent PDBselect 
{\small ($ftp.embl-heidelberg.de/pub/databases/protein\_extras/pdb\_select/recent.pdb\_select$)},

\noindent which contains proteins with
at most 25\% similarity between their sequences. We further selected from this
initial set those proteins that satisfy the following criteria: 
(i) The structures have very few missing atoms and/or amino acids; 
(ii) They have only a few amino acids with multiple coordinates; 
(iii) The secondary structure assignment must be in the
PDB header, and (iv) The data set should not include 
membrane proteins. With these criteria our 
database consists of
1177 proteins, containing both single and multi-chain proteins.
The shape parameters $\Delta$ and $S$ are calculated using the
coordinates of the proteins in the dataset.

{\bf Number of two- and three-body Contacts}

To understand the origin of the dense packing in proteins we have
enumerated the number of two and three body contacts. The nature
of these contacts is characterized by dividing the 20 amino 
acids into 9 classes based on their van der Waals volumes and
charge type (see for example Table 1.1 in ref 
\cite{Creightonbook93}):
(1) Gly; (2) Pro; (3) Cys; (4) small hydrophobs (sH): Ala, Val, Ile, Leu, Met;
(5) large hydrophobs (LH): Tyr, Trp, Phe; (6) small polar (sP): Ser, Thr;
(7) large polar (LP): Asn, Gln, His; (8) positively charged (+): Arg, Lys;
and (9) negatively charged (-): Asp, Glu. 

For each amino acid, we selected all its side-chain (s) heavy 
atoms to determine the s-s contacts. By representing
the backbone (b) by C$_{\alpha}$ atoms we also computed the number of 
b-b and s-b contacts. Two
side-chains are in contact if at least one pair of heavy atoms 
(from the two residues) is 
separated by at most 5.2 \AA. Two backbones are in
contact when the distance between them is at most 6.5 \AA. A 
side-chain
and a backbone make a contact when the minimal distance between a pair of
their atoms is at most 5.5 \AA. We also determined the various types of 
3-body contacts: (b,b,b), (b,b,s), (b,s,s) and (s,s,s) using the same
cutoff distances between the atoms of the three residues.\\

{\bf RESULTS}

{\bf Radius of gyration of folded state obeys Flory law}

\begin{figure}[ht]
\includegraphics[width=6.75in]{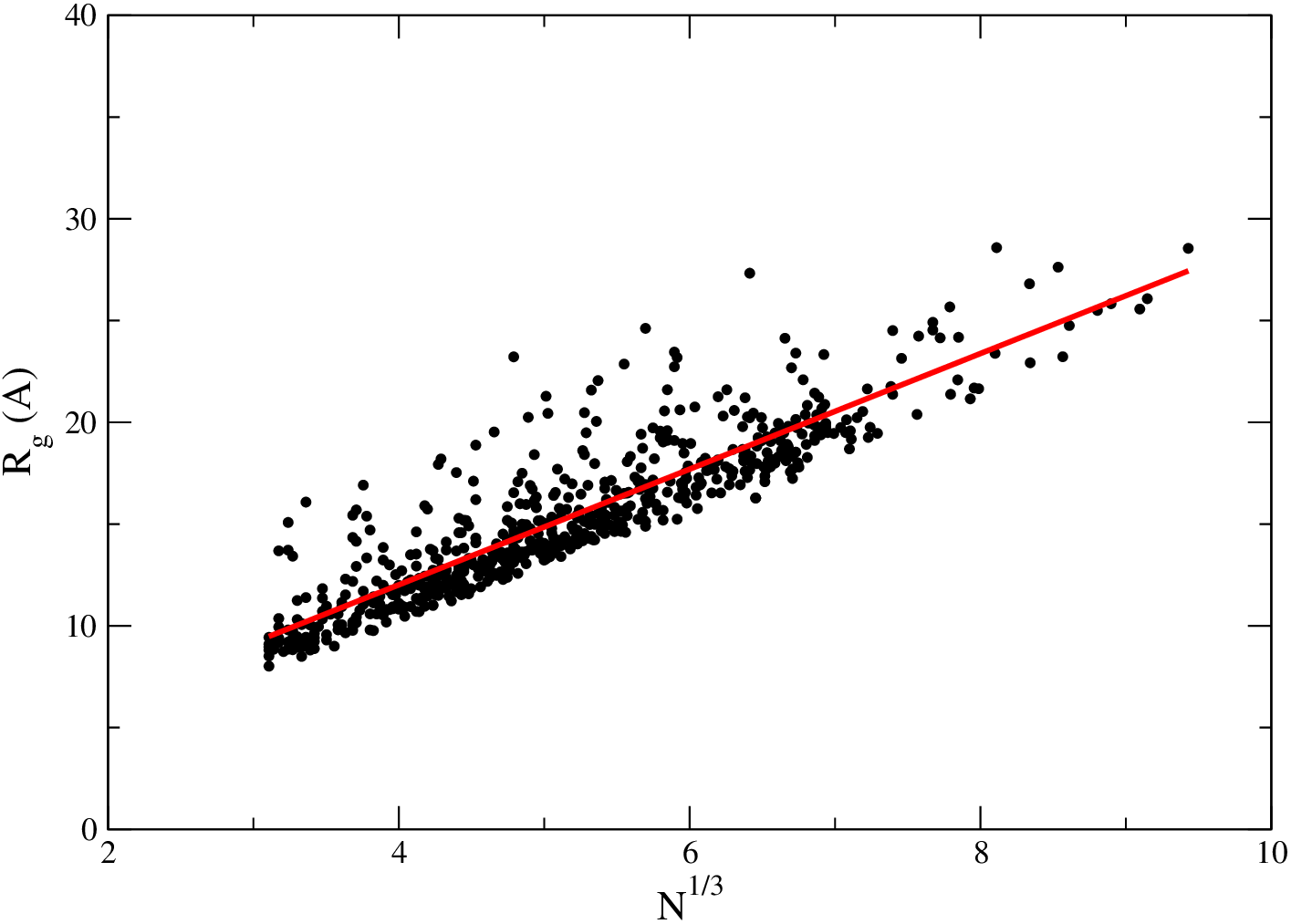}
\caption{\label{fig:Rg_one}}
\end{figure}

For a perfectly spherical molecule, volume, an extensive variable, is
$V = \frac{4\pi}{3} R_{g}^{3}$.
From this it follows that for a spherical molecule 
$R_g \sim a N^{1/3}$
where $N$ is the number of amino acid residues.
The plot of R$_g$, for the dataset of proteins, as a function of $N$ 
confirms the Flory result ($R_g \sim a N^{1/3}$ with $a = 3$ \AA) 
with a correlation of 0.90 (Fig.\ref{fig:Rg_one}).

Dobson and collaborators \cite{WilkinsBiochem99} measured, 
by pulse field gradient NMR techniques, the distribution 
of hydrodynamic radii for a set of natively folded proteins.
The data fits the empirical equation:
$R_h = (4.75 \pm 1.11) N^{0.29 \pm 0.02}$.
For a spherical chain, 
$R_g = \sqrt(\frac{3}{5}) R_h$ 
and therefore
\[
R_g \approx (3.67 \pm 0.86) N^{0.29 \pm 0.02}
\]
which is in agreement with the best fit to our data
(Fig.\ref{fig:Rg_one}). Because 
$R_{g} \sim 3 N^{1/3}$, as predicted by Flory for compact globules,
we expect that proteins should be spherical. 
From the values of $R_g$ alone it would seem that proteins are
maximally compact and are likely to produce close packed
structures. However, a detailed
examination of the shapes shows (see below) that this 
is not the case.

{\bf Single chain proteins are aspherical}

The distribution P($\Delta$) of $\Delta$ for the proteins in
PDBselect is peaked at $\Delta \approx$ 0.1 for single-chain proteins
(Fig.(\ref{fig:Asphericity_total})). Despite a relatively
long tail at large values of $\Delta$, about 78\% of the 
403 single-chain
proteins from PDBselect have asphericities smaller than 0.2. This 
shows that, to a first approximation, the native-state conformations 
of globular single-chain proteins can be modeled as spheres.
A similar behavior is found for chains that are part of multi-chain 
proteins, but in this case the distribution is considerably broader 
(see middle panel of Fig.(\ref{fig:Asphericity_total})).

\begin{figure}[ht]
\includegraphics[width=6.75in]{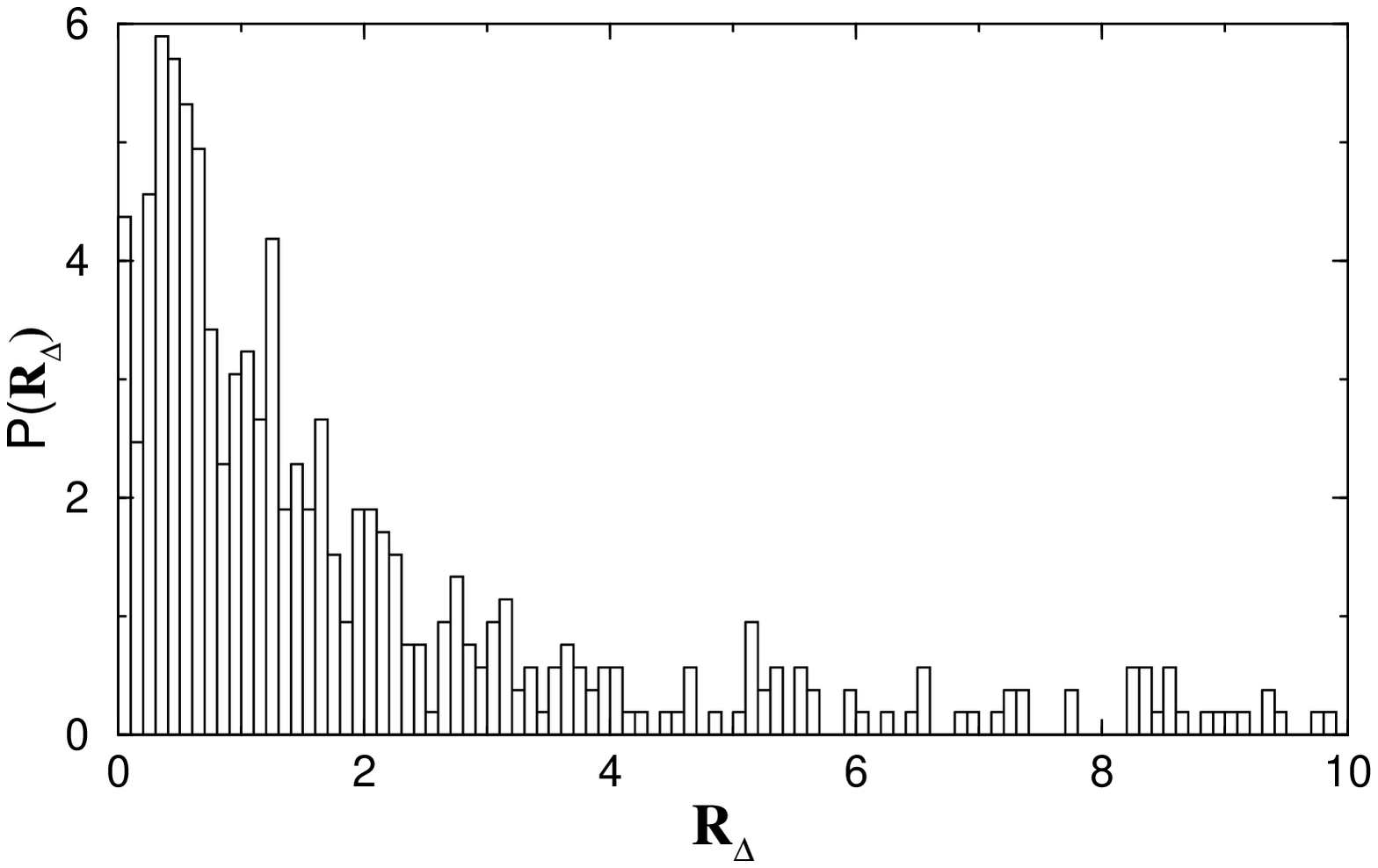}
\caption{\label{fig:ratio_Tot_av_Ad}}
\end{figure}

Many proteins are organized as oligomers. For oligomeric proteins 
a clear deviation from sphericity appears when considering the 
whole oligomers. For oligomers the distribution is
broad. Only about 50\% of oligomeric proteins have
$\Delta$ values smaller than 0.2 (bottom panel in
Fig.(\ref{fig:Asphericity_total})). 
The deviation from sphericity is also evident in the 
distribution P($R_{\Delta}$) where 
$R_{\Delta} = \frac{\Delta_{O}}{\Delta_{I}}$ with
$\Delta_{O}$ and $\Delta_{I}$ are the asphericity parameters for 
the oligomer and the individual chains in the oligomer
(Fig.(\ref{fig:ratio_Tot_av_Ad})). If both
the oligomer and the individual chains have similar values of
$\Delta$, we expect P($R_{\Delta}$) to be peaked around 1.
The distribution P($R_{\Delta}$) is peaked at
$R_{\Delta}$ between (0.4-0.6) (Fig.(\ref{fig:ratio_Tot_av_Ad})). 
The majority (60\%) of oligomers have $R_{\Delta} > $ 1. 
In other words the oligomer is less spherical than its chains. 
The results in Fig.(\ref{fig:Asphericity_total}) show that, 
in general, the approximation that isolated proteins are spheres
with $R_g \sim a N^{1/3}$, which is used for the interpretation
of a wide variety of experimental techniques
(like sedimentation analysis and gel filtration), is likely to 
produce small errors. However, such an approximation for oligomeric
proteins is inaccurate.

Although the bulk of globular proteins have relatively 
low values of
$\Delta$ ($<$ 0.2), there are a number of proteins in the tail 
of the distribution.
Two proteins for which both $\Delta_{O}$ and $\Delta_{I}$
have very large 
values ($>$ 0.80) are spectrin (PDB code 2spc) and a 
signaling protein (PDB code 1qu7). 
Both the individual chains and the whole 
protein are very elongated and the chains appear inter-twined 
in the oligomer. Large deviations from a spherical shape for 
these helical
proteins might be required for functional reasons. 

{\bf Majority of the proteins are prolate}

The distribution of the shape parameter for protein 
chains (see Fig.(\ref{fig:Shape_total})) reveals
that only about 20\% of 
single-chains have $S \approx 0$. Additional 20\% have  
$S <$ 0, that is they have an oblate shape, while the remaining 
60\% are prolate. 
For multi-chain proteins only about 10\% have $S \approx 0$. 
Even if $\Delta$ seems 
to indicate that usually proteins are approximately spherical, their actual 
shape tends to be prolate. Thus, there is considerable asymmetry 
in the shapes of globular proteins.

\begin{figure}[ht]
\includegraphics[width=6.00in]{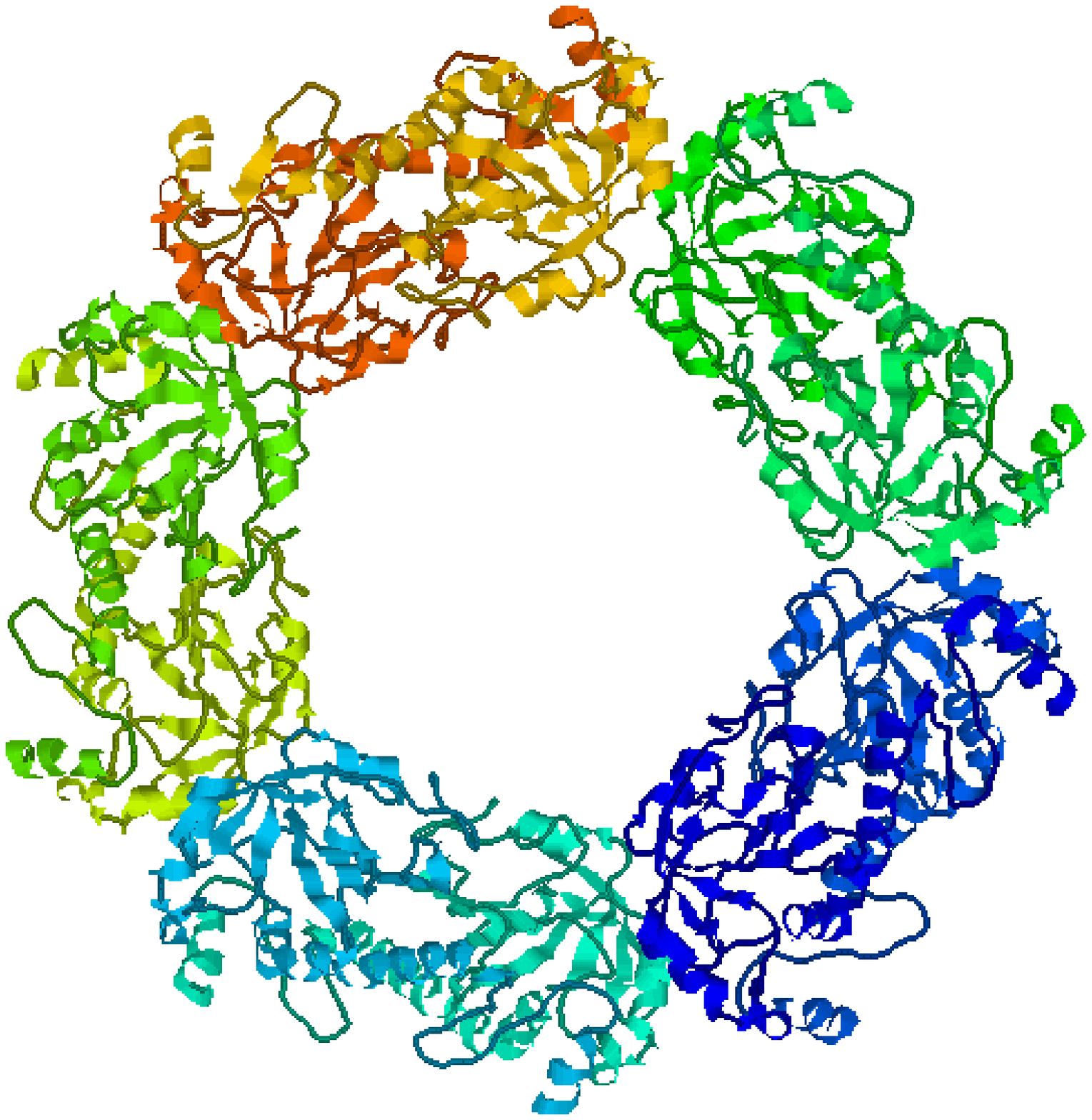}
\caption{\label{fig:rasmol_1qmv}}
\end{figure}

To put these findings into context we note that
the values of the shape parameter for real proteins are smaller 
than for Gaussian chains \cite{HoneycuttJCP89}. 
In the latter case $S >$ 0.63 which shows that Gaussian chains
are always prolate. Because Gaussian chains are useful models of
denatured states it is reasonable to assume that upon folding 
the chain
becomes compact but retains the prolate shape. It is therefore
surprising that about 20\% of proteins in the 
native state have oblate shapes, i.e. $S <$ 0.
Two such proteins with the largest (in absolute
value) negative $S$, and therefore having the most pronounced oblate 
shapes, 
are 1qb3 ($S$ = -0.164) and 1qmv ($S$ = -0.153). Both are multi-chain
proteins in which the identical individual chains are disposed in an 
almost perfect circular registry like in  
human peroxidase-B (1qmv) (Fig.(\ref{fig:rasmol_1qmv})). 

For proteins with $S <$ 0 there has to be substantial changes in the 
shape during the transition from denatured states to the
folded state. The helical proteins, 2spc and 1qu7, have 
the largest preference for prolate shape (with $S$ = 1.5 and 
1.8 respectively). More generally, the examination of
the distribution of $S$ and $\Delta$ for the 403 monomeric proteins 
shows that usually proteins with
small asphericities have also small values of $S$. 
This rule breaks down
somewhat unexpectedly for proteins with substantially 
oblate shapes (i.e., large negative $S$) which 
have relatively small asphericities especially when compared to 
the proteins 
with pronounced prolate shapes (i.e., large positive $S$).

{\bf High degree of packing in proteins correlates with substantial contacts involving backbone atoms}

\begin{figure}[htbp]
\begin{center}
	\vspace{-0.75cm}
	\includegraphics[width=6.5in]{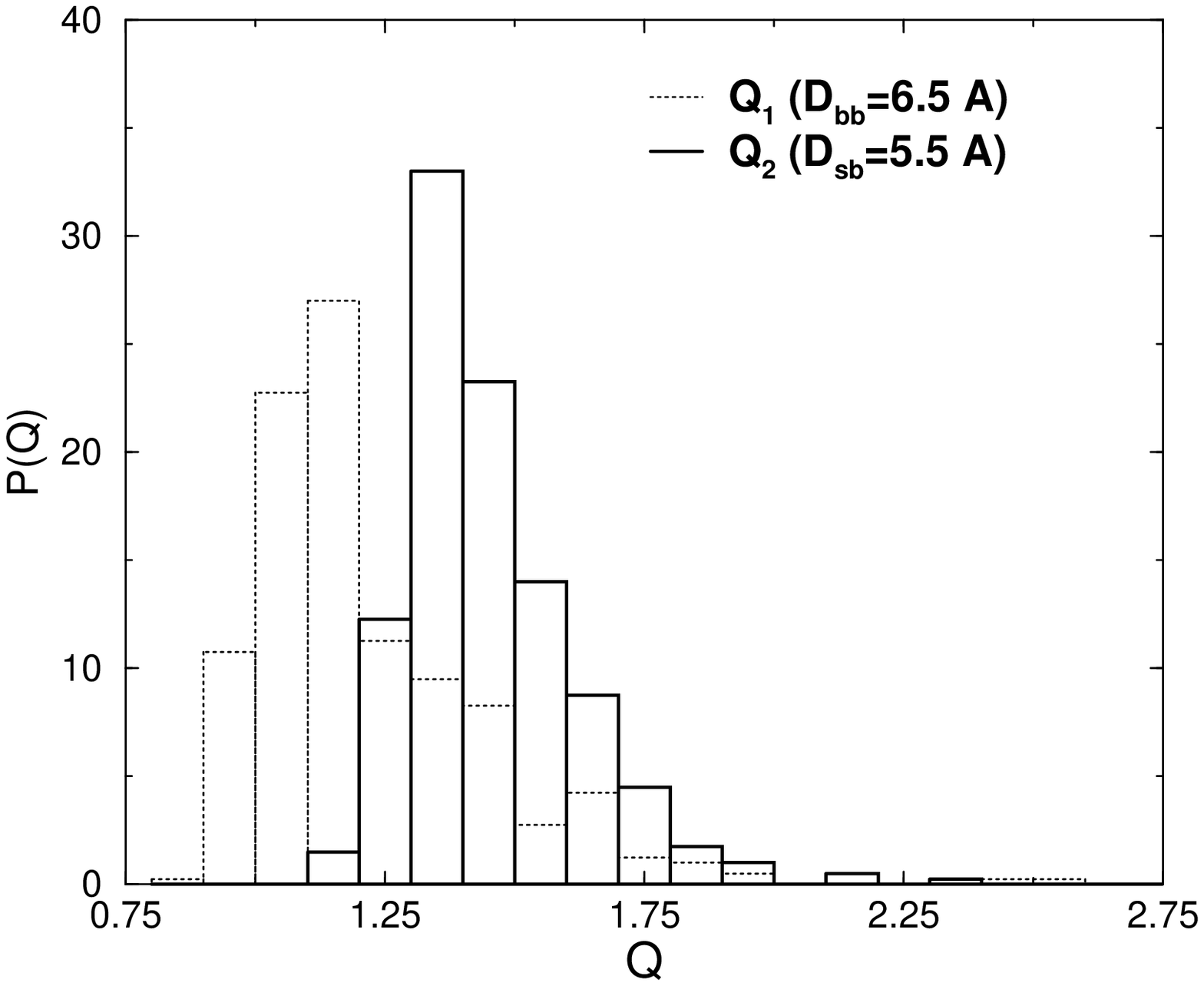}
\caption{}
\label{fig:ratio_2body_btot_ss.PDB_one}
\end{center}
\end{figure}

To probe the origin of the small values of $\Delta$, 
that is linked to high packing density in nearly 80\% of
globular proteins, we analyzed the number of two-body and three-body contacts 
in our dataset of proteins. We have previously reported that there are a 
large number of two-body contacts involving the protein backbone 
\cite{BetancourtJCP02} (N. V. Buchete, J. E. Straub and D.T., submitted to Protein Sci.).
Using the 9 classes of amino acids (see the
{\bf Methods} section) with the cut-offs $D_{bb}$ = 6.5 \AA,
$D_{sb}$ = 5.5 \AA, and $D_{ss}$ = 5.2 \AA, we determined the number 
and nature of two and three-body contact between backbones, between 
backbones and side-chains and between side-chains.
The distributions P($Q_{1}$) of $Q_{1} = \frac{N_{bb}}{N_{ss}}$ and P($Q_{2}$) of 
$Q_{2} = \frac{N_{sb}}{N_{ss}}$ where $N_{bb}$ is
the number of b-b contacts, and $N_{sb}$ is
the number of s-b contacts, and $N_{ss}$ is the 
number of s-s contacts in the
403 single-chain proteins 
(Fig.(\ref{fig:ratio_2body_btot_ss.PDB_one}))
have several interesting and somewhat unexpected features. The
distributions are peaked at values of $Q_1$ and $Q_2$  
larger than 1, the shift being most
evident for the s-b contacts. This implies that, 
in the native states of globular 
proteins, contacts between backbone atoms 
and especially those between backbones 
and side-chains occur at least as frequently as contacts between the 
side-chains. Secondly, both distributions have relatively
long tails that extend well into the region corresponding to 
$Q_1$ and $Q_2$ greater than 1.5. 
Upon closer inspection we find that in general the b-b and s-b 
contacts occur in a correlated manner with the s-s contacts. 
In other words, if
two residues are found to have their side chains in contact, 
then their backbones are also
very likely to be in contact or at least the backbone of one 
residue is in
contact to the side-chain of the other residue. In addition, the 
backbones of the 
pairs of their adjacent residues along the chain are also 
usually in contact with the other or/and with the side-chains. 
These distributions are similar for chains in
the multi-chain proteins from PDBselect (data not shown). 
The only difference is that in this case P($Q_1$)
is peaked at somewhat lower values while P($Q_2$)
is peaked at slightly larger values than in the case of
single-chain proteins. 

Analysis of the nature of amino acids that participate in the
s-s contacts shows that, as expected, 
the largest percentages come from contacts 
between hydrophobic residues. 
On an average 18\% are contacts between small
hydrophobs (Ala, Val, Ile, Leu, Met) and 12\% are between 
small hydrophobs 
and large hydrophobs (Tyr, Trp, Phe). Contacts among large 
hydrophobs are 
scarce as they represent only approximately 1.5\% of all s-s contacts.

Enumeration of the number of 3-body contacts shows
a pronounced shift with respect to the distributions 
of 2-body contacts. The number of
contacts between side-chains, which is on the 
the same order as the number of contacts between 2 side-chains 
and a backbone, is typically larger than contacts 
between backbones and between 2 backbones and a side-chain. 
The distribution of the types of s-s-s contacts shows 
that the two
strongest contributions are due to interactions among small 
hydrophobs or between 2 small hydrophobs and a large hydrophob. 
The contacts between three large hydrophobs become even less favored;
in about 80\% of proteins their contribution adds-up to less than
1\% of the total number of s-s-s contacts. Other types of 3-body 
contacts that appear as highly unfavored in the proteins from our 
dataset are contacts
between 3 Gly residues, 3 Cys residues, 3 Pro residues or between 
3 charged
amino acids of the same sign. Due to Coulomb repulsion, 
contacts between any other combination of 3 charged residues are 
infrequent,
usually contributing with about 0.4\% to the total number of s-s-s
contacts while the contribution from contacts between 3 charges 
of the same sign is at best 0.06\%. 
The presence of near equal number of 
two- and three-body contacts 
involving backbone and side chains confers tight packing which in 
turn explains the low $\Delta$ values in proteins.

{\bf Shape parameters for ``denatured'' states of proteins}

A complete understanding of how proteins fold requires 
characterizing the structures of the ensemble of unfolded states. 
Several
recent studies have argued that unfolded states can have
considerable residual structure even at elevated denaturant 
concentrations 
\cite{ShortleScience01,KleinSeetharamanScience02,MokJMB99,WongJMB96,SaabRinconBiochem96}. 
To a first approximation, unfolded states
can be modeled as random coils. 
Therefore, the radius of gyration for denatured states should be 
given by $R_g \approx l_{u} N^{\nu_{F}}$ where the Flory exponent
$\nu_F \approx$ 0.6 and $l_{u}$ is a ``persistence'' length.
The Gaussian chains (corresponding to $\Theta$-solvent condition)
are the simplest approximation for random coils. The values of
$\Delta$ and $S$ for Gaussian chains can serve as estimates of 
the shape parameters for unfolded states of proteins.
Explicit calculation of the 
asphericity parameter for long Gaussian chains, 
averaged over an ensemble of denatured states, 
shows that \cite{DiehlJPhysA89}

\begin{equation}
< \Delta > = \frac{1}{8} ( 26 - 9 K_3 )
\label{eqn:Delta_Gaussian_chain}
\end{equation}
where $K_3 = \int _{0} ^{\infty} dx x^{5/2} (sinh x) ^{-3/2}$. Thus,
numerical calculations give $\Delta$ = 0.52 and $S$ = 0.87. 
If excluded volume
interactions are taken into account we find that 
$\Delta$ = 0.55 and $S$ = 0.92. 
These calculations show
that, if denatured states are modeled as random coils
without side chains 
then the average asphericity should show
considerable deviation from globularity. It also
follows that large changes in $\Delta$ and $S$ are
accompanying the folding process.

Because interactions between side chains and backbone play 
a crucial role
in determining the packing in proteins realistic estimates of
$\Delta$ and $S$ even for denatured states must include
side chain effects. To obtain reliable values of 
$\Delta$ and $S$ for models of denatured states
we used Langevin simulations of an off-lattice model of 
the 34-residue WW-domain.
These calculations provide a picture of the states sampled by the
denatured state ensemble (DSE) in a good and a poor solvent 
conditions. Solvent quality is mimicked by tuning the
interaction potentials between side-chains (see below).
 
Each residue in a chain is represented by two beads: one is the
position of its C$_{\alpha}$ carbon atom 
and the other one is the center of mass
of its side-chain. The conformation of the chain is specified by 
the set of
C$_{\alpha}$ positions ($\overrightarrow{r_{b,i}}$) and the set 
of side-chain 
centers of mass positions ($\overrightarrow{r_{s,i}}$).
The potential energy of each conformation is given by
\begin{equation}
E_{pot} (\{\overrightarrow{r_{b,i}}\},\{\overrightarrow{r_{s,i}}\}) = V_{BL} + V_{SBC} + V_{BA} + V_{DIH} +V_{NB}
\label{eqn:pot_en_conf}
\end{equation}
where $V_{BL}$, $V_{SBC}$, $V_{BA}$, $V_{DIH}$ and $V_{NB}$ 
are bond-length
potential, side-chain to backbone connectivity potential, bond-angle
potential, dihedral angle potential and non-bonded long-range 
potential. The bond length potential is

\begin{equation}
V_{BL} = \sum_{i = 1}^{N-1} \frac{k_r}{2} (|\overrightarrow{r_{b,i+1}} - \overrightarrow{r_{b,i}}| - r^{WT}_{i,i+1})^{2}
\label{eqn:V_BL}
\end{equation}
where $k_r$ = 200 $\epsilon _h$/a$^2$, a = 3.8 \AA, $\epsilon _h$ 
is the average
strength of the hydrophobic interaction (i.e., it is the unit of 
energy in our
model) and $r^{WT}_{i,i+1}$ is the distance between the 
$C_{\alpha}$s at 
positions $i$ and $i+1$ in the wild-type sequence from the 
PDB (N-terminal of 1pin).

\begin{equation}
V_{SBC} = \sum_{i = 1}^{N} \frac{k_r}{2} (|\overrightarrow{r_{s,i}} - \overrightarrow{r_{b,i}}| - r^{WT}_{sb,i})^{2}
\label{eqn:V_SBC}
\end{equation}
where $r^{WT}_{sb,i}$ is the distance 
between the $C_{\alpha}$ and 
the center of mass of the side-chain at position $i$ in the wild-type 
sequence from the PDB, and $V_{BA}$ is given by

\begin{equation}
V_{BA} = \sum_{i = 1}^{N-2} \frac{k_{\theta}}{2} (\theta _i - \theta ^{WT}_i)^{2}
\label{eqn:V_BA}
\end{equation}
where $k_{\theta}$ = 20 $\epsilon _h$/(rad)$^2$ 
and $\theta^{WT}_{i}$ is the 
bond angle at positions $i$ in the wild-type sequence from the PDB.
The dihedral angle potential is given by

\begin{equation}
V_{DIH} = \sum_{i = 1}^{N-3} [A_i(1 - cosd\phi) + B_i(1 + cos 3d\phi) + C_i(1 - sind\phi)]
\label{eqn:V_DIH}
\end{equation}
where $A_i$ = 1.0$\epsilon _h$, $B_i$ = 1.6$\epsilon _h$, 
$C_i$ = 2.0$\epsilon_h$, $d\phi = \phi - (\phi^{WT}-60.5)$ 
when $\phi^{WT} < 2\pi/3$, $A_i$ = 1.2$\epsilon _h$, 
$B_i$ = 1.2$\epsilon _h$, 
$C_i$ = 0.0$\epsilon _h$, $d\phi = \phi - (\phi^{WT}-180)$ when 
$ 2\pi/3 < \phi ^{WT} < 4\pi/3$ and
$A_i$ = 1.0$\epsilon _h$, $B_i$ = 1.6$\epsilon _h$, 
$C_i$ = 2.0$\epsilon _h$, $d\phi = \phi - (\phi^{WT}-299.5)$ 
when $ 4\pi/3 < \phi ^{WT} < 2\pi$. The non-bonded potential
that gives rise to globularity is given by

\begin{equation}
V_{NB} = \sum_{i = 1}^{N-3} \sum_{j = i+3}^{N} V_{ij}(r_{ij})
\label{eqn:V_NB}
\end{equation}
where $V_{ij} (r_{ij}) = 4 \epsilon_h (\frac{R_{ij}}{r_{ij}})^{12}$ 
in good solvents and 
$V_{ij} (r_{ij}) = 4 \epsilon_h [(\frac{R_{ij}}{r_{ij}})^{12} - (\frac{R_{ij}}{r_{ij}})^6]$ for bad solvents. 
The values for $R_{ij}$ were 
taken from Table I in \cite{KolinskiJCP93}. 

It has been argued that the two major factors that contribute 
to packing in proteins are steric effects and energetic interactions 
between the largely hydrophobic core in the protein interior
\cite{RichardsQuartRevBioph94,KussellJMB01}.
Modeling energetic effects is more difficult than excluded volume
interactions. To dissect these two contributions separately, 
we first neglect attractive interactions between the side chain 
$C_{\beta}$ spheres. This corresponds to good solvent conditions.
In this case the unfolded state is expected to have the  characteristics
of Flory random coil. In our model the coil to globule transition is induced 
by assigning attractive interactions, that mimics poor solvent conditions, 
between the $C_{\beta}$ spheres. Assigning a uniform value for $\epsilon_h$,
which masks the sequence dependence entirely, allows us to examine
shapes of maximally compact globular structures.
Comparison of the $\Delta$ and $S$ values
for the globules, which are favored under poor solvent conditions,
and the native state gives an indication of the dispersion
in interactions that stabilizes a given fold.

We performed Langevin simulations in the underdamped limit with 
a low friction
coefficient, $\mu$ = 0.05$\tau_L^{-1}$, and a time integration step
$h$ = 0.005$\tau_L$, where 
$\tau_L$ = $(\frac{ma^2}{\epsilon_h})^{1/2}$ \cite{GuoBiopol95}. 
We generated 15 trajectories, each starting from different 
initial conditions. All trajectories start at a high temperature 
(T = 3.0 in units of $\frac{\epsilon_h}{k_B}$) 
and after a quench to T = 2.5 the temperature is
reduced sequentially in steps of 0.050 degrees until 
the final temperature, T = 1.0, is
reached. At each temperature the shape 
parameters are determined as averages over 0.5 $\times 10^6$ 
steps, but only 
after an initial lag of 10$^6$ steps which are needed for 
equilibration.

\begin{figure}[htbp]
	\vspace{-1.25cm}
	\hspace{-3.cm}
	\subfigure[]{
	\label{fig:WWdomain_noattract}
	\includegraphics[width=3.75in]{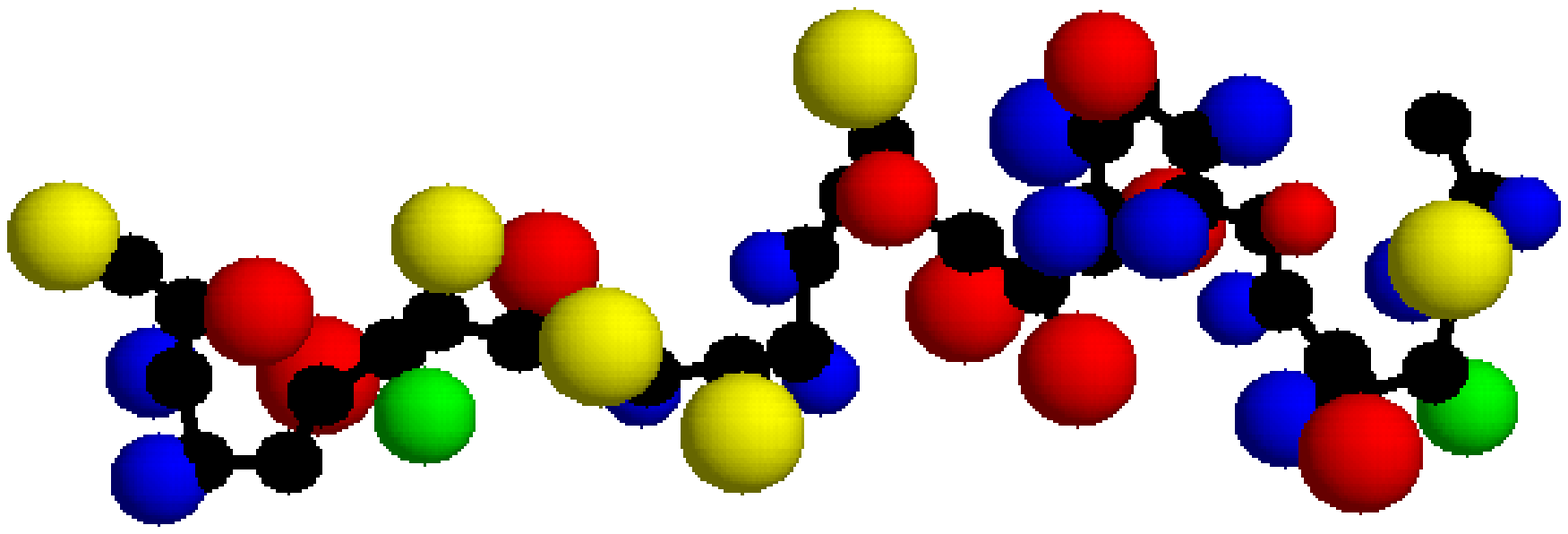}}
	\subfigure[]{
	\label{fig:WWdomain_attracton}
	\includegraphics[width=2.75in]{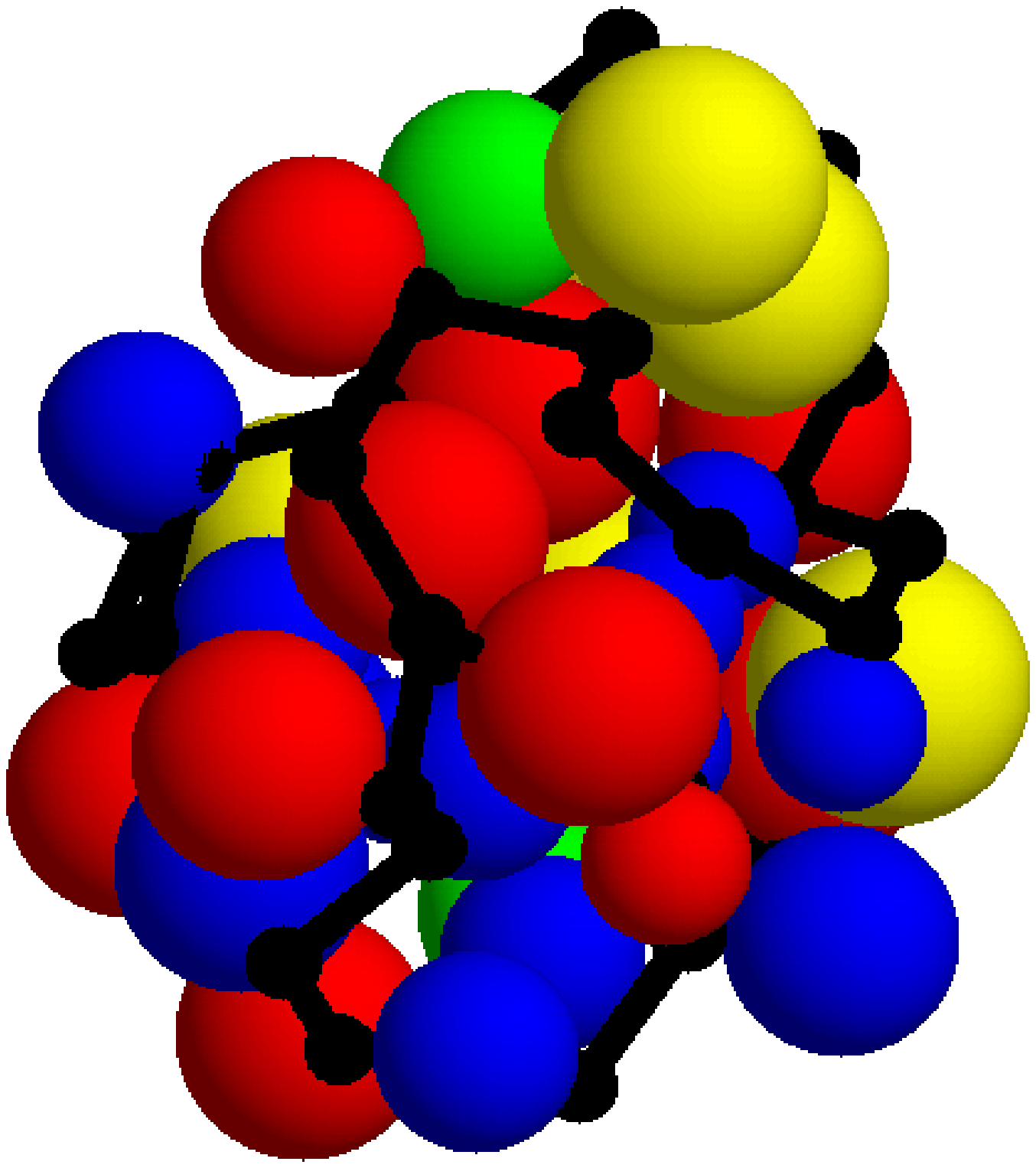}}
	\\
        \subfigure[]{
	\label{fig:WWdomain_PDB}
	\includegraphics[width=3.95in]{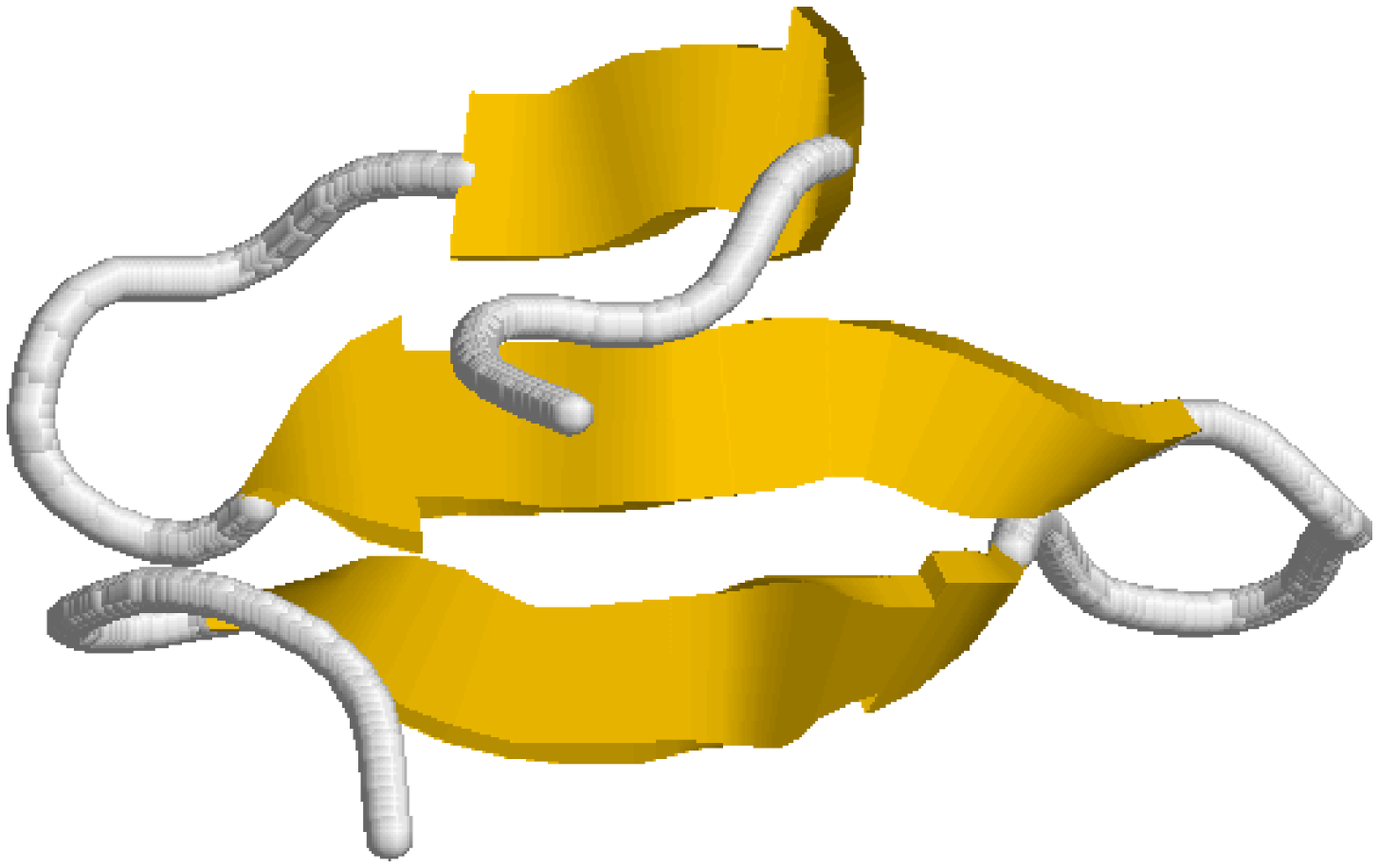}}
\caption{}
\label{fig:graphs_WWdomain}
\end{figure}

In good solvents $\Delta$ and $S$ for the WW-like domain
are 0.5 and 0.7
respectively. The $S$ values are less than those
obtained by neglecting side chains. A typical conformation 
of the Flory random coil is
shown in Fig.(\ref{fig:WWdomain_noattract}). 
When the solvent is poor, so that the coil to
globule transition occurs, $\Delta$ and $S$ reduce drastically.
For our model $\Delta$ = 0.04 and $S$ = 0.01 which implies that the
presence of side-chains drives the structures to acquire maximally compact
spherical shapes. (Fig.(\ref{fig:WWdomain_attracton})).
The spherical shape formed under poor solvent conditions with
$\Delta \approx S \approx$ 0 is different from the most probable
values of $\Delta$ and $S$ for proteins 
(Figs.(\ref{fig:Asphericity_total}) and (\ref{fig:Shape_total})). 
In our model the radii of side chains coincide with the
van der Waals sizes of the residues in the WW-domain. Thus, steric
effects which represent one factor in packing are properly taken
into account. However, hydrophobic interactions in our model are 
uniform ($\epsilon _{h}$ is the same for interactions 
between all side chains). This is not the case for the WW-domain
in which dispersion in hydrophobic interactions along 
with energetics associated with hydrogen bonds formation are necessary to
produce the three-stranded $\beta$-sheet protein.  In the
native state $\Delta$ and $S$ for the WW-domain are 0.15
and 0.07 respectively. These values, which are different from the
ones for the maximally compact structures (nearly close packed) 
(Fig.(\ref{fig:WWdomain_attracton})), show that both excluded
volume and energetic stabilization are important in providing the observed
packing in proteins.
These conclusions are consistent with the findings of 
Kussell et al. \cite{KussellJMB01} who showed that besides excluded volume interactions
energetic effects (hydrophobic, electrostatic, and polar) must play a role
in determining the fold of a protein.
\\

{\bf CONCLUSIONS}:

A hallmark of native states of proteins is that their cores are 
densely packed \cite{RichardsQuartRevBioph94}. 
In this paper we have shown that, although the radii of
gyration of folded proteins follow the Flory law 
$R_g \sim 3 N^{1/3}$, there is considerable 
asymmetry in the shapes of proteins.
Oligomeric proteins in general are found to deviate substantially 
from spherical shapes. 
The majority of single chain proteins are prolate ($S >$ 0) 
with small $\Delta$ values ($<$ 0.2). 
Large deviations from sphericity show
that proteins are not close packed, which is in accord with
the conclusions of \cite{LiangBiophysJ01}. This observation indicates
that a vast number of proteins can tolerate substantial volume
mutations. We hypothesize that in proteins with larger 
$\Delta$ and $S$
values mutations that change the van der Waals volume of side-chains
in the core can be made without compromising their stability or
the network of contacts.

Mutagenesis experiments on T4 lysozyme and $\lambda$-repressor 
lend support to our 
hypothesis. The $\Delta$ (0.20) and $S$ (0.19) values for the 
wild-type T4 lysozyme, a 164 residue single chain protein 
(3lzm), are larger than those for a
typical single-chain protein (Fig(\ref{fig:Asphericity_total})). 
Various volume mutations in the
hydrophobic core of T4 lysozyme
(32 of the approximately 350 structures of 
T4 lysozyme variants deposited in the PDB) lead to only minor
increases in the $\Delta$ and $S$ values (data not shown). 
The average $\Delta$ and $S$ for these mutants are 0.22 and 0.20 respectively.
Irrespective of the
sign of the volume change, the overwhelming majority of these mutations
lead to increased stability of the structure (see Table 3 in
\cite{RichardsQuartRevBioph94}). In addition, comparison 
of the number of
3-body s-s-s contacts in the WT T4 lysozyme and in one of the
mutants (140l), shows that only 10\% of the contacts are different.
This suggests that in proteins with large $\Delta$ and $S$
non-polar core mutations induce only minimal rearrangements in the
network of contacts. 

A corollary of our hypothesis is that in
fairly spherical proteins volume mutations would lead to a
larger degree of rearrangements in the network of contacts. 
Indeed, upon analysis of the WT N-term of the $\lambda$-repressor (1lbm) and
a mutant (1lli), both structures are
spherical ($\Delta \approx$ 0.08 and $S \approx$ 0.01) but
there is a 18\% difference between the number of 
3-body s-s-s contacts.
These two examples (mutational experiments on
T4 lysozyme and $\lambda$-repressor) provide
anecdotal evidence for our hypothesis that volume mutations in
proteins with large $\Delta$ and $S$ can be readily made without
drastically altering the number of contacts. On the other hand,
in highly spherical proteins substantial rearrangements of contacts 
accompanied perhaps by shifts in secondary structures occur
upon mutations that change the volume of the side-chains.

The dense packing in most proteins arises because there is a  
substantial number of contacts between side chains and backbone atoms.
A number of recent studies
\cite{BaharFoldDes96,MaritanNature00,KussellJMB01,BagciJCP02,BucheteJCP03,BagciProteins03} 
have also concluded that orientations of
side chains and their interactions with the protein backbone 
are important in determining the shapes of proteins.
In this paper we have shown that the total number of
b-b and s-b contacts is of the same order of magnitude as the number
of s-s contacts. This is also true at the level of 3-body contacts.
Our results also suggest that the knowledge-based
interaction potentials should include terms that account for the 
large number of b-b and s-b contacts \cite{BucheteJCP03}. 
Statistical potentials
that focus only on interaction parameters between side chains
cannot successfully account for native state stability.

A model of denatured states of proteins using realistic
representations ($C_{\alpha} - C_{\beta}$ model) shows that
unfolded states are more spherical than polymer models that neglect
side chains. Even when the solvent is good we find
that $\Delta$ and $S$ values are less than the theoretically 
predicted values for Gaussian chains or chains with excluded 
volume. The simulations of collapsed
structures for a model of WW-like domain that neglects the
dispersion in the hydrophobic interactions shows that
$\Delta \approx S \approx$ 0. These values, which are 
substantially different for the WT structure, show that
the variations in the sequence-dependent energies
and excluded volume interactions are both 
responsible for the high density packing observed in proteins.

{\bf Acknowledgments:} We are grateful to Dr. J. D. Honeycutt,
Dr. M. Molisana and Dr. Margaret S. Cheung for useful discussions. This work was
supported in part by a grant from the National Science Foundation
through grant number NSF CHE-03-02 9340.

\newpage

\begin{center}
{\bf Figure Captions}
\end{center}

Fig.\ref{fig:Rg_one}: Radius of gyration, R$_g$, as a function
of N, the number of amino acid residues for the 403 monomeric proteins 
in our dataset.
The best linear fit to the data, with a correlation 
of 0.9, gives $R_g = 3 N^{0.333}$.

Fig.\ref{fig:Asphericity_total}: Distribution of the asphericity 
parameter $\Delta$ for the proteins in the dataset. The top panel, 
that corresponds to P($\Delta$) for the single-chain proteins, 
shows that a substantial fraction have $\Delta <$ 0.1. The
middle panel shows P($\Delta$) for single chain proteins
that are part of oligomers. In contrast to the top panel,
P($\Delta$) is broad which implies that single chain proteins 
adopt aspherical shapes when they oligomerize. Large
deviations from sphericity are seen in multi chain proteins
(bottom panel).

Fig.\ref{fig:ratio_Tot_av_Ad}: Distribution of $R_{\Delta}$
for the 536 multi-chain proteins. Here, 
$R_{\Delta} = \frac{\Delta_O}{\Delta_I}$, where $\Delta_O$ is the
asphericity parameter for the oligomer and $\Delta_I$ is the
corresponding value for the individual chains. The fraction
$\int_{1}^{\infty} P(R_{\Delta}) dR_{\Delta} \approx$ 0.6 
which implies that oligomers are more asymmetric than the
individual chains from which they are constructed. 

Fig.\ref{fig:Shape_total}: Distribution of P($S$) of $S$, the shape 
parameter from (Eq.(\ref{eqn:Shape_def})). The distributions 
of P($S$) in the top and middle panels are similar which shows 
that the shapes of single chain proteins (either in isolation or as
part of a multi-chain) are roughly the same. The distribution
P($S$) for the oligomers are broader with a larger fraction
of $S <$ 0  corresponding to an overall oblate shape.

Fig.\ref{fig:rasmol_1qmv}: Structure of the 
human thioredoxin peroxidase-B (1qmv) which has
one of the most negative values of $S$ ($S$ = -0.15). 
In the denatured states it is likely that the overall
shape is prolate. Thus, during folding and oligomerization
there is a qualitative and large change in the overall shapes
of such proteins.

Fig.\ref{fig:ratio_2body_btot_ss.PDB_one}: Distributions of 
$Q_{1} = \frac{N_{bb}}{N_{ss}}$ and $Q_{2} = \frac{N_{sb}}{N_{ss}}$
where $N_{bb}$ is the number of b-b contacts, $N_{sb}$ is
the number of s-b contacts, and $N_{ss}$ is the 
number of s-s contacts in the 403 single-chain proteins. 
Both distributions are peaked at values larger than 1. 
This shows that in the native states of globular 
proteins contacts involving backbone atoms and especially 
those between backbones and side-chains occur frequently.

Fig.\ref{fig:WWdomain_noattract}: Typical conformation of the 
off-lattice model $C_{\alpha}-C_{\beta}$ 
of the WW-like domain in a good solvent is a coil.
The shape parameters for the Flory coil $\Delta$ and $S$ are 0.5 and 0.7, 
respectively. These
are much larger than the values for the native states of
proteins seen in Figs.(\ref{fig:Asphericity_total}) and
(\ref{fig:Shape_total}). (b): Typical globular conformation of the 
off-lattice model of the WW-like domain in a bad solvent.
In this case $\Delta$ = 0.04 and $S$ = 0.01 which implies that the
presence of the attractive interactions between side-chains  
drives the structures to acquire a spherical shape.
(c): Conformation of the WT WW-domain (1pin). The $\Delta$ and $S$
values are 0.15 and 0.07 respectively. Comparison with (b) shows
that the acquisition of the three stranded $\beta$-sheet occurs at
the expense of deviation from maximally compact structures. It is
likely that WT WW-domain can tolerate volume mutations without 
compromising its stability.

\newpage

\[
\]

\newpage

\begin{figure}[ht]
\[
\]
\includegraphics[width=7.25in]{Asphericity_total.eps}
\caption{\label{fig:Asphericity_total}}
\end{figure}
\[
\]

\newpage

\[
\]

\newpage

\begin{figure}[ht]
\includegraphics[width=7.25in]{Shape_total.eps}
\caption{\label{fig:Shape_total}}
\end{figure}
\[
\]

\newpage

\[
\]

\newpage

\end{document}